\documentclass{iopart}
\usepackage{graphicx}
\begin{document}
\title{Self-consistent study of multipole response in neutron rich nuclei using a modified realistic  potential}
\author{D Bianco$^{1,2}$\footnote{present address : Ecole Normale Superieur (ENS) de Cachan, 61 av. du Pr\'esident Wilson, Cachan - France} ,
F Knapp$^{3}$,  N Lo Iudice$^{1,2}$, P Vesel\'{y}$^{2}$, F Andreozzi$^{1,2}$, G De Gregorio$^{1,2}$, A Porrino$^{1,2}$   }
\address{$^1$Dipartimento di Fisica, Universit$\grave{\rm{a}}$ di Napoli Federico II, Monte S. Angelo, via Cintia, I-80126 Napoli, Italy
 }
 \address{$^2$ Istituto Nazionale di Fisica Nucleare (INFN), Sezione di Napoli,
 Monte S. Angelo, via Cintia, I-80126 Napoli, Italy}
\address{$^3$ Faculty of Mathematics and Physics, Charles University, Prague, Czech Republic}
\ead{loiudice@na.infn.it}

%\date{\today}
\begin{abstract}
The   multipole response of  neutron rich O and Sn isotopes is computed in  Tamm-Dancoff and  random-phase approximations using the canonical Hartree-Fock-Bogoliubov  quasi-particle basis.  The calculations are performed using an intrinsic Hamiltonian composed of a $V_{lowk}$   potential, deduced from the CD-Bonn nucleon-nucleon interaction, corrected with phenomenological density dependent and  spin-orbit terms. The effect of these two pieces on energies and multipole responses is discussed. The problem of removing the spurious admixtures induced by the center of mass motion and by the violation of the number of particles is investigated. The differences between the two theoretical approaches are discussed quantitatively. Attention is then focused on the dipole strength distribution, including  the low-lying transitions associated to the pygmy resonance. Monopole and quadrupole responses are also briefly investigated. A detailed comparison with the available experimental spectra contributes to clarify the extent of validity of the two self-consistent approaches.  
\end{abstract}
%\pacs{21.60.Ev,21.10.Re,24.30.Cz}
\maketitle
 
\section{Introduction} 
During the last decades,   the  quasi-particle  random-phase approximation (QRPA)  calculations using a  self-consistent Hartree-Fock-Bogoliubov (HFB) basis were rooted in the density functional theory. The     energy density functionals   were usually  derived    from Skyrme  
\cite{Doba84,Doba96,engel99,Khan02,Tera05,Yosh08,Miz09,Losa} or Gogny forces \cite{Giamb03,Peru08} and from solving relativistic mean field equations \cite{Paar03}. A more exhaustive list of references can be found in Refs.\cite{Bend03,Vret05}. These approaches   were adopted with success to  study  bulk properties of nuclei and nuclear responses of different multipolarities \cite{Tera05,Paar03}. 
 
An alternative route based on the use of realistic nucleon-nucleon (NN) potentials was attempted recently by Roth and coworkers \cite{Herg09,HergPap11}. They
applied first \cite{Herg09} a fully self-consistent HFB formalism \cite{RiSchu} to  an intrinsic Hamiltonian using a correlated UCOM effective interaction \cite{UCOM}  derived from the Argonne V18 interaction \cite{argonne}.

In a second step \cite{HergPap11}, the same authors  adopted a HFB canonical basis   to solve the eigenvalue problem within the framework of QRPA. They used new effective interactions derived from the same Argonne V18 NN potential   by means of the similarity renormalization group (SRG) approach \cite{SRG} and added  a phenomenological density dependent term simulating a three-body contact interaction.  This phenomenological term was of crucial importance for generating single particle spectra in qualitative agreement with the empirical ones.

They showed that  the self-consistent QRPA calculation   yields a complete decoupling of the center of mass (CM) and number operator  spurious states  if the single particle states are expanded in a sufficiently large harmonic oscillator space.  Moreover, their systematic study of  the nuclear response for various multipolarities  has  produced some encouraging results.
The   description of the giant resonances resulted to be of quality comparable to other more phenomenological QRPA studies. Moreover, it was possible to associate the low-lying transitions  to the Pygmy modes.
 
Several deviations from experiments still remain. One source of these discrepancies is the insufficient spin-orbit splitting resulting from the HFB solutions. 

As in Ref. \cite{Herg09}, we have generated a canonical HFB quasi-particle (qp) basis. We started with using  a $V_{lowk}$ potential   derived from the CD-Bonn NN interaction \cite{Mcl}. 
Such a potential was introduced more than a decade ago \cite{Bogner01}. It smooths out the original $NN$ potential   and, therefore, is suited for HF and HFB calculations. Its intimate connection with the renormalization group approaches was pointed out \cite{Bogner03}. The potential has been used extensively in shell model spectroscopic studies \cite{covrev}. 

We, then,   added  to $V_{lowk}$ the same phenomenological density dependent potential used  in Ref. \cite{HergPap11} in order to reduce the too large spacing between qp levels produced by the $V_{lowk}$ potential.

We further included  a phenomenological spin-orbit term. The latter piece is meant to enhance the splitting between spin-orbit partners and, thereby, obtain a better detailed agreement with the empirical single particle spectra.

The canonical HFB basis was then adopted to solve the eigenvalue problem in QRPA and within an upgraded qp Tamm-Dancoff approximation (QTDA). This upgraded QTDA adopts  a  basis of states orthogonalized by Gramm-Schmidt to the CM and number operator spurious states within the two quasi-particle space.      Although more involved than standard QTDA, it allows to eliminate exactly the spurious admixtures induced by either the CM motion or by the violation of the number of particles, without being forced to use spaces of huge dimensions. 

The comparative analysis of QTDA and QRPA calculations gives us the opportunity of establishing quantitatively the detailed differences between  the two approaches. We shall see that they yield very close results  in several cases.

Moreover, the QTDA calculations are   relevant to our future studies. In fact, the  QTDA phonons are the building blocks of the multiphonon states obtained in the equation of motion phonon method (EMPM) \cite{Bianco} which represents an alternative to the many approaches developed recently for extending the QRPA \cite{Paar07,LoI12}. The EMPM  was already employed with success in the study of the low energy $E1$ spectrum in $^{208}$Pb \cite{Biancoa}. Thus, generating the QTDA phonons in a HFB basis represents the first step for a fully self-consistent calculation to be carried out within the EMPM.

We will consider different multipole responses   within both  QTDA and QRPA for sets of neutron rich O and Sn isotopes.  We  will concentrate mainly 
on the structure of the giant dipole resonance (GDR) and pay attention also   at the low-energy dipole transitions near the neutron decay threshold.
These   excitations have been extensively studied in the lasts decades and tentatively interpreted as manifestation of  the pygmy dipole resonance (PDR) promoted by a translational oscillation of the  neutron skin against  a $N=Z$ core

The first tentative experimental evidence of the PDR was gained  in Coulomb excitation experiments  on neutron rich oxygen isotopes   \cite{Leiste,tryg02,tryg03}. The nature of the observed low-lying peaks was also   investigated theoretically in   shell model \cite{Saga} as well as in a QRPA plus phonon-coupling model \cite{CoBo01}.

The low-lying dipole excitations were investigated extensively also in Sn isotopes in several experiments \cite{Adrich05,Oze07,Endres10} and several theoretical approaches \cite{Sar04,Tzon08,Avdee11,Klim07,Daut11}. A review may be found in Refs. \cite{Paar07,Aumann08}.

It is to be pointed out that the purpose of the present work is  not to compete with the approaches based on energy density functionals, but rather  to try to identify the phenomenological terms which are to be added to an effective potential derived from a realistic $NN$ interaction in order to obtain a  description of the nuclear response in fair agreement with experiments.
The identification of these terms may give useful hints on how   realistic NN potentials should be modified in order to be usefully spent for reliable studies of the nuclear response.  The ultimate goal is, in fact, to perform  nuclear spectroscopy calculations based on the exclusive use of these potentials.

\section{Hartree-Fock-Bogoliubov }

The HFB  transformation is defined \cite{RiSchu} as
\begin{eqnarray}
\label{HFBtr}
\beta^{\dagger}_{r}=\sum_{s} \Bigl(U_{sr}c^{\dagger}_{s}+V_{sr}c_{s}\Bigr), \nonumber\\
\beta_{r}=\sum_{s} \Bigl(U_{sr}^{*}c_{s}+V_{sr}^{*}c^{\dagger}_{s}\Bigr).
\end{eqnarray} 
It transforms the  creation and annihilation particle operators $c_s^\dagger$ and $c_s$, respectively, into the corresponding quasi-particle operators $\beta^{\dagger}_{r}$ and $\beta_{r}$.
Its coefficients fulfill the  conditions
\begin{eqnarray} 
\label{UUVV}
&&U^{\dagger}U+V^{\dagger}V=I, \qquad UU^{\dagger}+V^{*}V^T=I, \\
&&U^{T}V+V^{T}U=0, \qquad UV^{\dagger}+ V^{*}U^{T}=0,
\label{UVUV}
\end{eqnarray}
which ensure that the anticommutation relations are preserved.

The HFB ground state   defines (up to a unitary transformation) the qp vacuum $\mid 0 \rangle$.
Let us consider a Hamiltonian composed of a kinetic term $T$ and a two-body potential $V$. Its ground state expectation value   
\begin{equation}
E_{0} =\langle 0 \mid H \mid 0 \rangle  = \langle 0 \mid (T + V) \mid 0 \rangle  
\end{equation}
 is a   functional of the the density matrix   and the pairing tensor defined respectively as
\begin{eqnarray} 
\label{rho} 
 \rho_{rs}= \langle 0 \mid c^{\dagger}_{s}c_{r} \mid 0 \rangle = (V^{*}V^{T} )_{rs},\\
 \kappa_{rs}= \langle 0 \mid c_{s}c_{r} \mid 0 \rangle = (V^{*}U^{T} )_{rs}.
 \label{kappa}
  \end{eqnarray}
By performing a variation of   $E_{0}$ with respect to  $\rho$ and   $\kappa$, under the constraint   
\begin{equation}
\label{ncons}  
tr \rho =   N    
  \end{equation}
ensuring that the number $N$ of particles is conserved on average,   
one obtains the HFB eigenvalue equations
\begin{eqnarray}
\label{HFBmat}
\left( \begin{array}{cc}
h-\lambda &\Delta \\ 
- \Delta^*&-h^* +\lambda 
 \end{array} \right)
\left( \begin{array}{c}
U \\ V
\end{array} \right)=E_{qp} \left( \begin{array}{c}
 U \\ V
 \end{array} \right),
\end{eqnarray}
where
\begin{eqnarray} 
h_{rs} =t_{rs} +\sum_{tq}   V_{rtsq}  \rho_{qt}, \\
\Delta_{rs}=\frac{1}{2}\sum_{qt}V_{srtq}\kappa_{qt}.
\end{eqnarray}
These are a set of non linear equations   to be solved self-consistently.  
  
For practical purposes it is convenient to adopt the canonical basis. As shown   in Ref. \cite{RiSchu}, in such a basis
the density matrix $\rho$ is diagonal with eigenvalues $v^2_r$, while   the coefficient $u_r^2$ is deduced from  $v_r^2$ through the
normalization conditions (\ref{UUVV}) which become
\begin{equation}
\label{norm}
 u^{2}_{r}+v^{2}_{r}=1.
 \end{equation}
One can define the qp energy 
\begin{equation}
E_r = \sqrt{(\epsilon_{r}-\lambda)^{2}+  \Delta_{r}^2},
  \end{equation}
  where $\epsilon_{r} =h_{rr}$ and $\Delta_{r}= \Delta_{r \bar{r}}$.
The  chemical potential $\lambda$ is fixed by the number conserving condition (\ref{ncons}), which in the canonical basis becomes
\begin{equation}
\sum_{r}v^{2}_{r}=N.     
\end{equation}
The Bogoliubov coefficients assume the BCS-like expressions
\begin{eqnarray}     
u^{2}_{r}=\frac{1}{2} \left(1 + \frac{\epsilon_{r}-\lambda}{E_r} \right), \quad \quad \quad
v^{2}_{r}=\frac{1}{2 }\left(1 -  \frac{\epsilon_{r}-\lambda}{E_r} \right).
\end{eqnarray}   
It is to be pointed out that the energies $E_r$  are the diagonal matrix elements of the HFB Hamiltonian in the canonical basis. They do not coincide in general with  the qp HFB eigenvalues $E_{qp}$ obtained from solving the HFB equations (\ref{HFBmat}).  
 
\section{QTDA and QRPA in the HFB canonical basis}

\subsection{The quasi-particle Hamiltonian}
When  expressed in terms of the canonical qp operators $\alpha^{\dagger}_{r}$ and  $\bar{\alpha}_{r} =  (-)^{j_r + m_r} \alpha_{r j_r - m_r}$, the starting Hamiltonian becomes
\begin{equation}
H = E_0 + H_{11} + {\cal V},
\end{equation}
where $E_0$ is the HFB  ground state energy,  $H_{11}$ is a one-body qp Hamiltonian, and ${\cal V}$   a two-body potential describing the interaction among  quasi-particles.  
The one-body piece in the angular momentum coupled scheme has the expression
\begin{eqnarray} 
\label{Hamqp}
H_{11}=\sum_{rs} [r]^{1/2} {\cal E}_{rs} [\alpha^{\dagger}_{r} \times \bar{\alpha}_{s}]^0, 
\end{eqnarray}
where  $[r] = 2 j_r + 1$ and  the symbol $\times$ denotes coupling of two tensor operators to angular momentum $\Omega$. The other quantity is
\begin{equation}
\label{H11}
{\cal E}_{rs} = (\epsilon_{rs} - \lambda \delta_{rs}) (u_r u_s - v_r v_s) + \Delta_{rs} (u_r v_s + v_r u_s),
\end{equation}
where
\begin{eqnarray}
 \epsilon_{rs} = t_{rs}    +  \Gamma_{rs }
\end{eqnarray}
and
\begin{eqnarray}
\label{Gamrs}
\Gamma_{rs}   =\frac{1}{[r]^{1/2}} \sum_{t} [t]^{1/2} F^0_{rs tt}v_{t}^2,\\
\Delta_{rs}   = \frac{1}{2} \frac{1}{[r]^{1/2}} \sum_t [t]^{1/2} V^0_{rstt} u_t v_t
\label{Delrs}
\end{eqnarray}
are the Hartree-Fock and pairing potentials, respectively.
We have introduced the quantity
\begin{eqnarray}
 F^\nu_{rt sq} =  \sum_\Omega (-)^{r +q - \nu -\Omega} W(rstq; \Omega \nu) V^\Omega_{rstq},
\end{eqnarray}
where $W$ is a Racah coefficient.

It is to be pointed out that the one-body qp Hamiltonian does not have the standard diagonal structure as it would  have been the case, had we used the full HFB basis.

The quasi-particle two-body piece can be written in the synthetic form 
\begin{eqnarray} 
\label{Vqp}
{\cal V} =-\frac{1}{4}\sum_{rstq \Omega }[\Omega]^{1/2} V^\Omega_{rstq}: \Bigl[(c^{\dagger}_{r} \times c^{\dagger}_{s})^\Omega \times (\bar{c}_{t} \times \bar{c}_{q})^\Omega\Bigr]^0:, 
\end{eqnarray}
where  $V^\Omega_{rstq} $  are   unnormalized antisymmetric two-body matrix elements and $: \dots :$ denotes normal order with respect to the HFB qp vacuum.

\subsection{QRPA and QTDA eigenvalue equations}
The QRPA creation operator    is
\begin{eqnarray}
O^\dagger_\lambda &=& \sum_{r \leq s} \Bigl[Y^\lambda_{rs} {\cal A}^\dagger_{rs \lambda} + Z^\lambda_{rs} \bar{{\cal A}}_{rs \lambda} \Bigr], 
\label{QPRAop}
\end{eqnarray}
where
\begin{eqnarray}
{\cal A}^\dagger_{rs\lambda} =   \zeta_{rs}   \Bigl(\alpha^\dagger_r \times \alpha^\dagger_s\Bigr)^\lambda, \quad \quad \quad
  \bar{{\cal A}}^\lambda_{rs} =   - \zeta_{rs}\Bigl(\bar{\alpha}_r \times \bar{\alpha}_s \Bigr)^\lambda, 
\end{eqnarray}
with $\zeta_{ab} = (1 + \delta_{ab})^{-1/2}$.
The QTDA operator is obtained by putting $Z=0$.

The QRPA eigenvalue equations are derived by standard techniques
\begin{eqnarray}
\label{QRPA}
\left( \begin{array}{cc}
A^\lambda & B^\lambda \\ 
- B^{\lambda*}&-A^{\lambda*}
 \end{array} \right)
\left( \begin{array}{c}
Y^\lambda \\Z^\lambda 
\end{array} \right)=\omega_\lambda \left( \begin{array}{c}
 Y^\lambda \\Z^\lambda
 \end{array} \right),
\end{eqnarray}
where $\omega_\lambda = E_\lambda - E_0$.
The block-matrices are defined as  
\begin{eqnarray}
A^\lambda_{rstq} =\langle 0 \mid [{\cal A}_{r s \lambda},[H, {\cal A}^\dagger_{t q \lambda}]] \mid 0 \rangle, \\
B^\lambda_{rstq} = \langle 0 \mid [{\cal A}_{r s \lambda},[H, {\cal A}_{t q \lambda}]] \mid 0 \rangle.
\end{eqnarray}
The block-diagonal matrix $A$ is the QTDA Hamiltonian matrix.
It is composed of two pieces
\begin{equation}
A^\lambda_{rs tq} =\zeta_{rs} \zeta_{tq} \Bigl[\langle (r \times s)^\lambda \mid H_{11} \mid (t \times q)^\lambda \rangle   
+ \langle (r \times s)^\lambda \mid {\cal V} \mid (t \times q)^\lambda \rangle \Bigr]. 
\end{equation}
The first piece is
\begin{equation}
 \langle (r \times s)^\lambda \mid H_{11} \mid (t \times q)^\lambda \rangle
 %\nonumber\\  
=  \delta_{sq}   {\cal E}_{rt}  
+ \delta_{rt} {\cal E}_{sq}  
- (-)^{r + s - \lambda} \Bigl[\delta_{st} {\cal E}_{rq}   
+ \delta_{rq}  {\cal E}_{st}  \Bigr].
\end{equation}
It is to be noticed that the above matrix element is non diagonal as it would be the case if computed in the  HFB basis.
 
The second term is the standard QTDA two-body matrix element
 \begin{eqnarray}
 \langle (r \times s)^\lambda \mid {\cal V} \mid (t \times q)^\lambda \rangle =
 %\nonumber\\ &&
 V^{\lambda}_{rstq} \Bigl( u_r u_s u_t u_q + v_r v_s v_t v_q \Bigr) + \\
 F^{\lambda}_{rstq} \Bigl( u_r v_s v_t u_q + v_r u_s u_t v_q \Bigr)
-  (-)^{t+q-\lambda} 
 F^{\lambda}_{rsqt} \Bigl( u_r v_s u_t v_q + v_r u_s v_t u_q \Bigr).\nonumber 
 \label{Vqptautau}
\end{eqnarray} 
The off-diagonal block, entering the QRPA matrix only, is given by 
 \begin{eqnarray}
B^{\lambda}_{rs tq} =      \zeta_{rs}  \zeta_{tq} \Bigl[ V^\lambda_{rstq} \bigl( u_r u_s v_t v_q + u_t u_q v_r v_s \bigr) \\
-F^\lambda_{r s t q}  \bigl( u_r v_s u_t v_q + u_s v_r u_q v_t \bigr) 
 + (-)^{t + q - \lambda}   F^\lambda_{rs qt}   \bigl( u_r v_s u_q v_t + u_s v_r u_t v_q \bigr) \Bigr].\nonumber 
\end{eqnarray}
The  expression of the $A$ matrix written above is valid for standard QTDA. As we shall see in 
Sect. \ref{spurious},  $A$  becomes more involved in our upgraded  QTDA approach, in which the eigenvalue problem is formulated within a space spanned by states, free of spurious admixtures, which are linear combinations of two quasi-particle states. Consequently,   also the two quasi-particle expansion coefficients of the eigenfunctions of this modified $A$ matrix have a more complex structure with respect to  the standard QTDA wavefunctions.

\subsection{Transition amplitudes of multipole operators}
The transition amplitude for a generic multipole operator is 
\begin{eqnarray}
< \lambda \parallel  {\cal  M}_\lambda \parallel  0  >   
  &=&\sum_{r \leq s}  <r \parallel {\cal M}_\lambda  \parallel s>  \zeta_{rs}\nonumber\\
 && \Bigl[u_r v_s + (-)^\lambda u_s v_r \Bigr] \Bigl[Y^{( \lambda)*}_{rs} - (-)^\lambda Z^{( \lambda)*}_{rs} \Bigr].  
\end{eqnarray} 
The  standard $E\lambda$   multipole operator  is 
\begin{eqnarray}
{\cal M} (E \lambda \mu) = \frac{e}{2}
\sum_{i=1}^A (1-\tau_3^i ) r_i^{\lambda} Y_{\lambda \mu}(\hat{r_i})
= {\cal M}_0 (E \lambda \mu)  + {\cal M}_1 (E \lambda \mu), \label{MElamb}
\end{eqnarray}
where $\tau_3 = 1$ for neutrons and $\tau_3 = -1$  for protons and ${\cal M}_\tau (E \lambda \mu)$ are the isoscalar ($\tau=0$) and isovector ($\tau =1$) components.

In order to remove partially the spurious admixtures induced by the center of mass (CM) motion we will use the modified isovector $E1$ operator
\begin{eqnarray}
 \label{E1effch}
{\cal M} (E 1 \mu) = e \frac{N}{A}
\sum_{p=1}^Z     r_p  Y_{1 \mu}(\hat{r_p})  
- e \frac{Z}{A}
\sum_{n=1}^N     r_n  Y_{1 \mu}(\hat{r_n}) 
\end{eqnarray}
obtained by subtracting the contribution of the CM operator.
This modification is needed only for the QRPA. As we shall see, it is  irrelevant to the our version of QTDA whose states are completely free of spurious admixtures.

The strength function is 
\begin{eqnarray}
{\cal S}(E\lambda, \omega) = \sum_{\nu }B_{\nu}  ( E\lambda )\,\delta (\omega -\omega_{\nu })
%\nonumber\\
\approx  
\sum_{\nu }B_{\nu}  ( E\lambda )\,\rho _{\Delta }(\omega -\omega_{\nu }).
\label{Strength}
\end{eqnarray}
Here  $\omega$ is the energy variable,  $\omega_\nu$ the energy of the transition of multipolarity $E\lambda$   from the ground to the $\nu_{th}$ excited state   of spin $J = \lambda$ and
\begin{equation}
\rho _{\Delta }(\omega - \omega_\nu)=\frac{\Delta}{2\pi } \frac{1}{(\omega - 
\omega_\nu)^{2}+(\frac{\Delta}{2})^{2}}  \label{weight}
\end{equation}
is  a Lorentzian  of width $\Delta$, which replaces the $\delta$ function as a
weight of  the reduced strength
\begin{equation}
B_{\nu} (E \lambda )=  
 \left| \langle  \nu, \lambda  \parallel {\cal M}(E\lambda)   \parallel
 0\rangle \right|^{2}.    \label{BE}
\end{equation}

\section{Calculations} 
\label{calc}
\subsection{Choice of the Hamiltonian}
We consider an intrinsic  Hamiltonian obtained by subtracting the CM kinetic energy $T_{CM}$ from  the shell model kinetic operator. 
The  new kinetic term is therefore
\begin{equation} 
\label{Tint}
T_{int} =  \frac{1}{2m} \sum_i p_i^2  - T_{CM} = T   + T_{2},
\end{equation}
where  
\begin{equation}
T = (1 -\frac{1}{A}) \frac{1}{2m} \sum_i p_i^2  
\end{equation}
is  a modified one-body kinetic term  and  
\begin{equation}
T_{2} =- \frac{1}{2m A} \sum_{i \neq j} \vec{p}_i \cdot \vec{p}_j
\end{equation}
is a two-body piece which will be incorporated into the  potential  $V$.
The full Hamiltonian is
\begin{equation}
H= T + V,
\end{equation}
where
\begin{equation}
V =   V_{so} + V_{2} + V_\rho.
\end{equation}
Here
\begin{equation}
V_2 =     V_{lowk} + T_2
\end{equation}
is a two-body potential composed of the CM  kinetic piece $T_2$ and   a  $V_{lowk}$  potential \cite{Bogner01} deduced from the NN CD-Bonn force \cite{Mcl} with a cut-off $\Lambda= 1.9$ $fm^{-1}$.
We have generated its matrix elements in the harmonic oscillator basis using the code of Ref. \cite{Enge}. 
 
In addition  to  $V_{2}$, the  potential $V$ includes two phenomenological pieces, a spin-orbit term 
\begin{equation}
V_{so} = C_{so} \sum_i \vec{l}_i \cdot \vec{s}_i  
\end{equation}
plus a density dependent two-body potential
\begin{equation}  
V_{\rho} = \sum_{i < j} v_\rho (ij),
\end{equation}
where
\begin{eqnarray}
v_\rho  = \frac{C_\rho}{6} (1 + P_\sigma) \rho(\frac{\vec{r}_1 + \vec{r}_2   }{2}) \delta(\vec{r}_1 - \vec{r}_2).
 \end{eqnarray}
 This   potential was introduced in Ref. \cite{HergPap11} and was shown  \cite{Waroq} to give to the ground state energy the same contribution of the contact  three-body interaction  
\begin{equation}
v_3 = C_\rho  \delta(\vec{r}_1 - \vec{r}_2)  \delta(\vec{r}_2 - \vec{r}_3).
\label{V-3body}
\end{equation}
Such a  three-body contact force    was shown to improve the description of bulk properties in closed shell nuclei
within a HF plus perturbation theory approach \cite{Gunter}.  

The parameters of the phenomenological density-dependent and spin-orbit pieces were determined  by a comparative analysis of the data in $^{16}$O and $^{132}$Sn. The parameter of the density dependent potential was such   as to reproduce on average the peaks of the giant dipole resonance (GDR) over the sets of oxygen and tin isotopes. We obtained $C_\rho = 2600$ MeV $fm^{6}$  for oxygen  and $C_\rho = 4200$ MeV $fm^{6}$ for tin.
The spin-orbit parameter was tuned so as to approach the empirical neutron spin-orbit separations given in Table II of Ref. \cite{HergPap11}, $\Delta \epsilon_{so}= 6.18$ MeV for $^{16}$O and $\Delta \epsilon_{so}= 6.53$ MeV for $^{132}$Sn.  We obtained   $C_{so}= -0.9$ MeV for oxygen  and   and $C_{so}= -0.6$ MeV for tin isotopes.

\subsection{Effect of the phenomenological Hamiltonian pieces}
The selective effects of the different pieces of the Hamiltonian are investigated for  $^{16}$O and $^{132}$Sn. In these doubly magic nuclei, HFB turns into HF and, therefore, yields single particle energies directly comparable to the empirical ones. This is not the case for the canonical basis in open shell nuclei.  

\begin{figure}[ht]
\includegraphics[width=30pc]{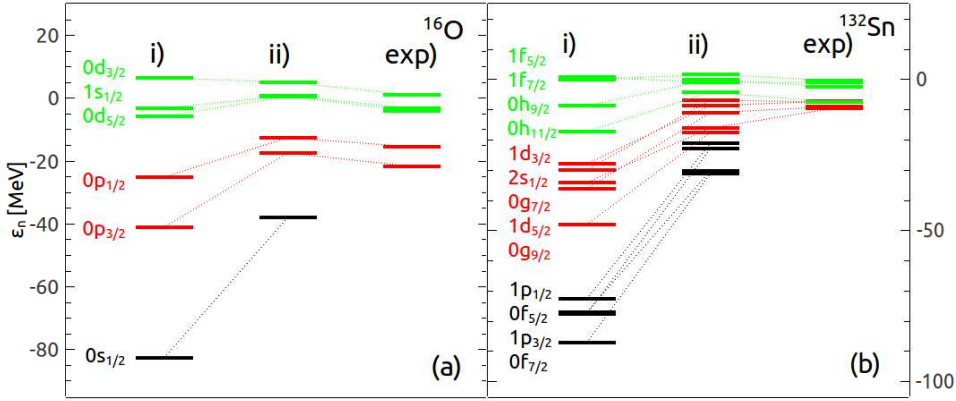}
\caption{  \label{fig1} 
Neutron single particle spectra in $^{16}$O (a)  and $^{132}$Sn (b)   with i)    $V_{2}$, ii) $V_{2} + V_\rho$. The empirical (exp) single particle levels are taken from \cite{Isak02}}
\end{figure}

\begin{figure}[ht]
\includegraphics[width=30pc]{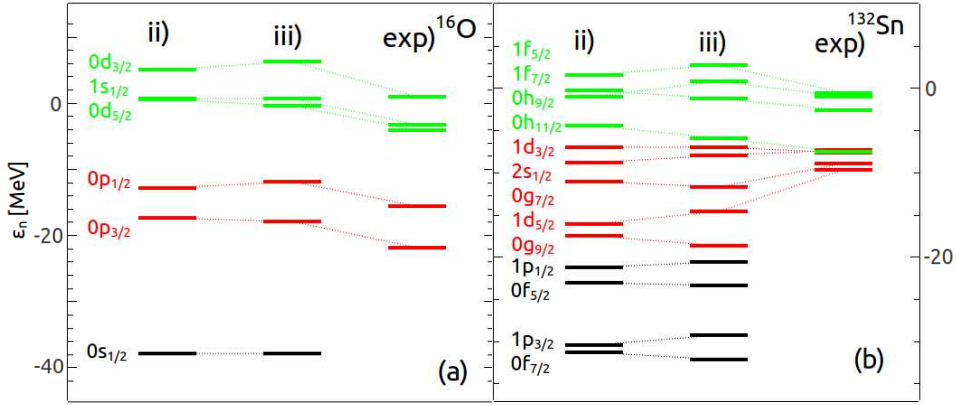}
\caption{  \label{fig2} 
Neutron single particle spectra in $^{16}$O (a)  and $^{132}$Sn (b)   with ii)  $V_{2} + V_\rho$, iii)$V_{2} + V_\rho + V_{so}$.
The empirical single particle levels  are taken from \cite{Isak02}} 
\end{figure}

As illustrated  in Figures \ref{fig1} and \ref{fig2},  the density dependent potential $V_{\rho}$ produces an overall drastic compression of the HFB spectrum resulting from  the use of $V_{2}$ only, in line with the calculation of Ref. \cite{HergPap11}.

The spin-orbit piece, by increasing the splitting between spin-orbit partners, has a fine tuning effect. It  pushes down in energy the spin-orbit intruders, thereby enlarging the gap between major shells,   and  compresses further the levels within each major shell,  consistently with the empirical single particle spectra. 

In tin isotopes, however, the levels within a major shell are not sufficiently packed as the empirical spectra would require. This  shortcoming does not allow an accurate detailed description of the positive parity low-lying QTDA or QRPA spectra. Thus, we will focus on
the dipole transitions and on the global features of monopole and quadrupole responses.

\begin{figure}[ht]
\includegraphics[width=30pc]{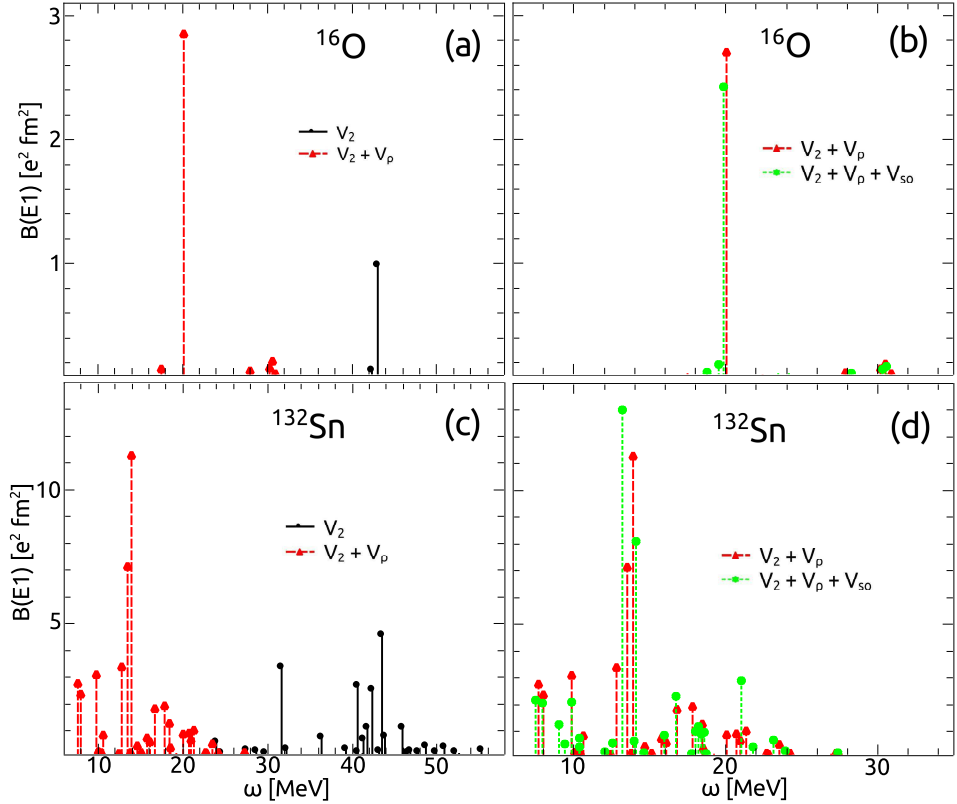}
\caption{  \label{fig3}   RPA $E1$ strengths   in  $^{16}$O ((a)  and (b)) and $^{132}$Sn ((c) and (d)) with   $V_{2}$ only and $V_{2} + V_{\rho}$ ((a) and (c)), with $V_{2} + V_{\rho}$ and $V_{2} + V_{\rho} + V_{so}$ ((b) and (d)). The centroids deduced from the experiments for $^{16}$O and $^{132}$Sn are, respectively,  $\sim 20.7$ MeV and  $\sim 16.1$ MeV  with corresponding widths $\sim 5.4$ MeV and   $\sim 4.7$ MeV.   
} 
\end{figure}

The  effect of the reduced level spacing, induced mainly   by $V_{\rho}$,,  gets especially manifest in the QRPA $E1$ spectra.   The transition amplitudes were computed using the $E1$ operator (\ref{E1effch}) which partly removes the spurious CM contribution. As shown in Figures \ref{fig3}, the  strength is concentrated into an energy interval which is about 20 MeV above the observed peaks, when   the $V_{lowk}$ plus the two-body kinetic term $T_2$ are used. It is shifted   downward to the experimental region by the density dependent potential $V_\rho$.   The spin-orbit piece induces only a slight redistribution of the strength.

An identical result is obtained in QTDA. As we shall see later, the $E1$   spectra hardly change in going from  QTDA to QRPA.

\begin{figure}[ht]
\includegraphics[width=30pc]{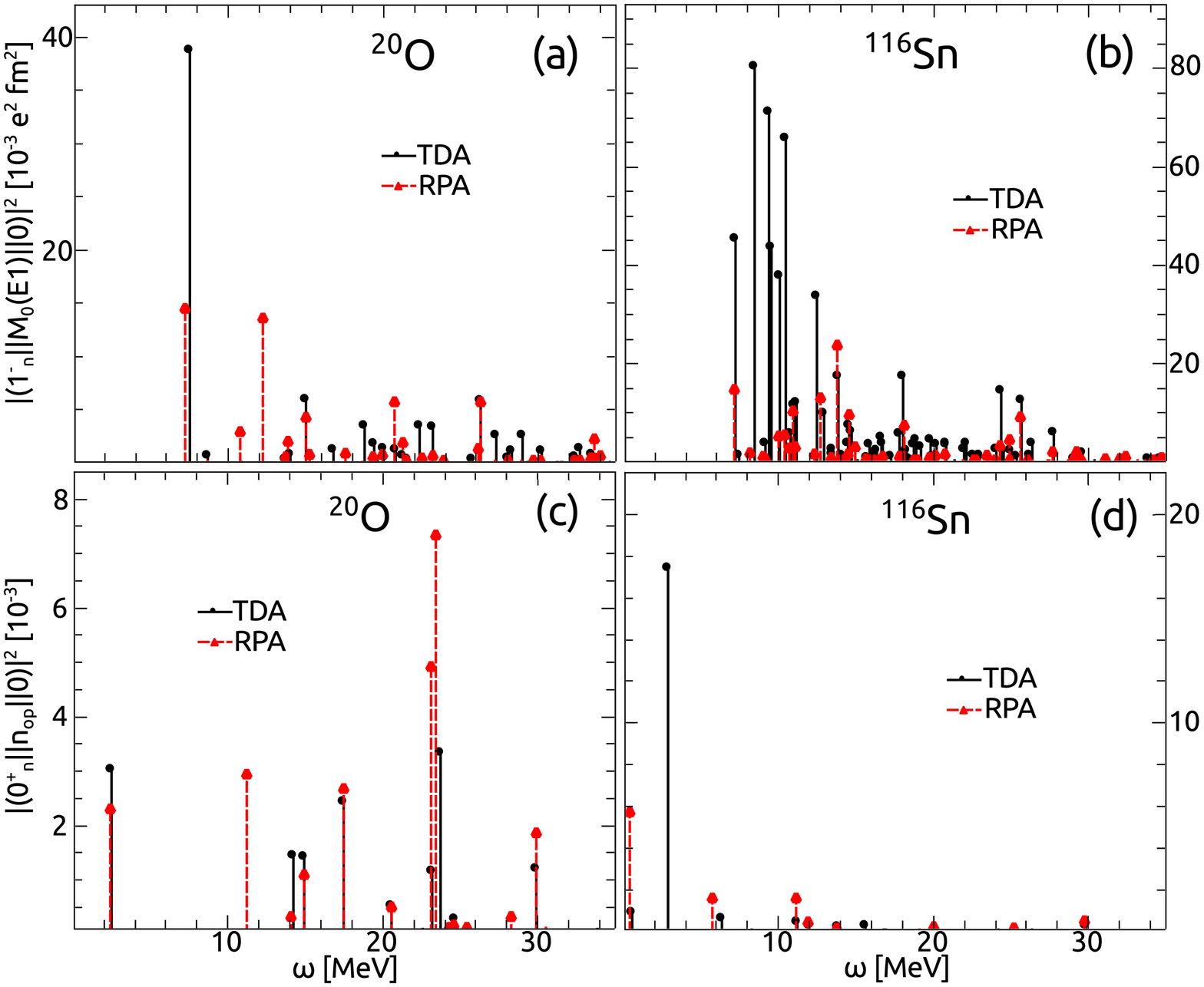}
\caption{  \label{fig4}  Strengths   of the transitions induced by the CM (a) and (b) and by number operators (c) and (d) in  $^{20}$O  and $^{116}$Sn.   
} 
\end{figure}

\subsection{Spurious admixtures}
\label{spurious}
In  a fully self-consistent  QRPA, the  $1^-$ and $0^+$ spurious states  lie at zero excitation energy and   collect the total    strength  induced by the CM and  the number operators, respectively.  

Numerically, their complete decoupling  from the physical intrinsic states is achieved only if a sufficiently large configuration space is adopted. This was the case of Ref. \cite{HergPap11}, where 15 major oscillator shells were considered in order to generate the HFB basis.

Having adopted a smaller space, which  includes up to 12 major oscillator shells, we do not achieve such a complete separation.   The energies of the lowest    $1^-$ and $0^+$ QRPA spurious states are   imaginary with   $\Im{(\omega_{1^-_1})}=0.18$ MeV and $\Im{(\omega_{0^+_1})}=1.64$ MeV respectively.  We should therefore expect some spurious admixtures, especially for the   QRPA $0^+$. 

The strength distributions of the isoscalar (CM)  $ {\cal M}_0 (E1\mu)$ (Eq. (\ref{MElamb})) and of  the number   $n_{op}$ operators  are shown in   Figs. \ref{fig4}. The isoscalar $E1$ strength is negligible for most but not all the $1^-$ states. Some of them, especially at low energy,  collect  up to $\sim 10^{-2}$ $e^2 fm^2$. The strength of the number operator is spread over the whole $0^+$ spectrum in $^{20}$O. It is concentrated at low energy in $^{116}$Sn.

Thus, we do not obtain a complete decoupling of the spurious $0^+$ and $1^-$ states as achieved in Ref. \cite{HergPap11}. The main reason is to be found in the more restricted harmonic oscillator space we    adopted (12   major shells versus 15). It is, however, to be pointed out that  a different potential was used  and a  different treatment  of its short range repulsion was made in Ref. \cite{HergPap11}. This difference might affect the efficiency in the procedure of eliminating the spurious admixtures. In any case, the use of effective charge in the $E1$ operator ( Eq. (\ref{E1effch}) ) removes partly the spurious CM strength.

A qualitatively similar result is obtained for standard QTDA.  The QTDA energy of the lowest $1^-$ is slightly negative. We get for instance $\omega_{CM} = -0.18$ MeV  in $^{20}$O and $\omega_{CM} = -0.640 $ MeV in $^{116}$Sn. In both nuclei, the state collects $\sim 98 - 99 \%$ of the CM coordinate strength. The residual strength contaminating the physical state is $\sim 1.8 \%$ in $^{20}$O and $\sim 1.5 \%$ in $^{116}$Sn. This spurious strength is distributed mainly among low energy states, as illustrated    in the same Fig. \ref{fig4}. Some of the low-lying QTDA states get an isoscalar $E1$ strength which is about three times larger than the corresponding QRPA states. Thus, although small, the residual spurious strength might alter the low-energy $E1$ spectrum.

The contamination of the $0^+$ spectrum is stronger. The energy of the lowest $0^+$ is $\omega_{0^+_1} =-1.226$ MeV in $^{20}$O and  $\omega_{0^+_1} =-0.429 $ MeV in $^{116}$Sn. The   number operator strength collected by the lowest $0^+_1$ is $\sim 95 \%$ in  $^{20}$O and $\sim 0.89 \%$ in $^{116}$Sn. The residual spurious strength is therefore larger,  $\sim 5.4 \%$ in $^{20}$O and $\sim 11 \%$ in $^{116}$Sn. It spreads over the whole spectrum in  $^{20}$O. It is instead concentrated mainly in a low energy peak in  $^{116}$Sn.  

Fortunately, in our improved version of QTDA,  we eliminate completely and exactly the spurious admixtures by applying  the  Gramm-Schmidt orthogonalization procedure to the two quasi-particle basis states with respect to the CM and number operator spurious states.
 
The spurious state, expanded in these two quasi-particle basis states, which we denote by $\mid i \rangle$, has the expression  
\begin{equation}
\mid \Phi_0 \rangle = \frac{1}{N(1)} \sum_{i=1}^n C_i \mid i \rangle, 
\end{equation}
where    
\begin{equation}
 N^2(1) =  \sum_{i = 1}^{n} |C_i|^2  
\end{equation}
is the normalization constant. The orthogonalized states have the expression
\begin{eqnarray}
\label{phik}
\mid \Phi_{k-1} \rangle = \frac{1}{N(k-1) N(k)} \Bigl [ N^2(k-1) \mid k-1 \rangle 
%\nonumber\\
- \sum_{i=k,n}C_{k-1}  C_i \mid i \rangle \Bigr], 
\end{eqnarray}
where $(k = 2, 3, ..n)$
\begin{equation}
 N^2(k) =  \sum_{i = k,n} |C_i|^2.   
\end{equation}
For $k=n$ the sum  $\sum_{i=k,n}$  disappears. So we have simply
\begin{eqnarray}
\label{phin}
\mid \Phi_{n-1} \rangle = 
%\nonumber\\
\frac{1}{N(n-1) N(n)} \Bigl [ N^2(n-1) \mid n-1 \rangle -  C_{n-1}  C_n \mid n \rangle \Bigr], 
\end{eqnarray}
where  
\begin{equation}
 N^2(n) =   |C_n|^2.   
\end{equation}
The CM spurious state $\Bigl(\lambda_1= (\kappa_1, 1^-) \Bigr)$ is 
\begin{eqnarray}
\mid \lambda_1 \rangle =   \frac{1}{N_1} R_{\mu} \mid 0 \rangle, 
\end{eqnarray}
where $R_{\mu}$ is the CM coordinate and $N_1$ the normalization constant.
Expanded in the two quasi-particle basis states, it acquires  the structure 
\begin{eqnarray}
\mid \lambda_1 \rangle =   \frac{1}{N_1} R_{\mu} \mid 0 \rangle=  \frac{1}{N_1} \sum_{r \leq s}   C^{\lambda_1}_{rs } \mid (r\times s)^{1^-} \rangle, 
\end{eqnarray}
where $C^{\lambda_1}_{rs }$ are the unnormalized coefficients  
\begin{eqnarray}
C^{\lambda_1}_{rs } = \sqrt{\frac{4 \pi}{9} } \frac{1}{A}\langle r \parallel r Y_1 \parallel s \rangle  ( u_r v_s - u_s v_r)
\end{eqnarray}
and   the normalization factor is given by
\begin{eqnarray}
N^2_1 =  \sum_{r \leq s}  |C^{\lambda_1}_{rs }|^2. 
\end{eqnarray}
Similarly, the number operator spurious state   ($\lambda_0= (\kappa_0, 0^+)$) is obtained by applying the number operator in normal order to the HFB vacuum. We get
\begin{eqnarray}
\mid \lambda_0 \rangle =  \frac{1}{N_0} \sum_{r }  C^{\lambda_0}_{rr }  \mid (r\times s)^{0^+} \rangle, 
\label{nopsp} 
\end{eqnarray}
where $C^{\lambda_0}_{rr } (\nu)$ are the unnormalized coefficients  
\begin{eqnarray}
C^{\lambda_0}_{rr }  =  \sqrt{2[r]} (u_r v_r) 
\end{eqnarray}
and $N_0$ is the normalization factor
\begin{eqnarray}
N^2_0 =  \sum_{r}  |C^{\lambda_0}_{rr } |^2.
\end{eqnarray}
The basis states obtained by the just outlined procedure are adopted to construct the Hamiltonian matrix  $\{\langle  \Phi_{r}\mid H \mid \Phi_s\rangle\}$. Its diagonalization  yields  eigenstates   rigorously free of spurious admixtures either induced by the CM excitation or by the violation of the particle number. The price we pay is that the Hamiltonian matrix  has a more involved structure and the eigenstates are given in term of the orthogonalized states  $\mid \Phi_r\rangle$.   Their QTDA standard structure  is finally recovered by using Eqs. (\ref{phik}) and (\ref{phin}) to express the states  $\mid \Phi_r\rangle$ in terms of  the two quasi-particle states.

The orthogonalization procedure  eliminates the negative energy spurious $1^-$ state resulting from the diagonalization of  the Hamiltonian  in the space spanned by the unmodified two quasi-particle basis states.

\begin{figure*}[ht]
\includegraphics[width=30pc]{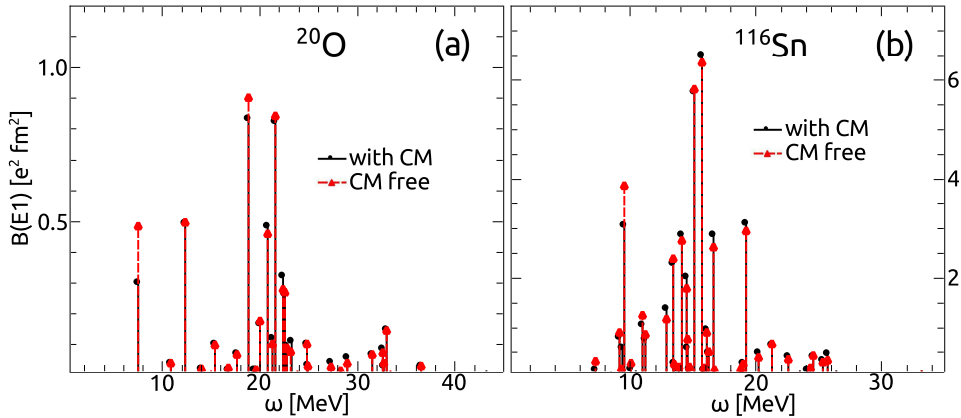}
\caption{  \label{fig5}   Strength distribution of the $E1$   transitions  obtained by using QTDA states with and without spurious admixtures, in  $^{20}$O (a) and $^{116}$Sn (b).  
} 
\end{figure*}

It has the additional important effect of removing the residual spuriousness from the remaining physical states, thereby modifying the $E1$ strength distribution. Figure \ref{fig5}   shows that the changes are very modest.  One can hardly notice any difference  between the spectra obtained with and without spurious  admixtures.
A small discrepancy is observed in the strong low-energy transitions.  Removing the spurious admixtures has the effect of enhancing their strength and, thereby, approaching   the  QRPA strength.

\begin{figure}[ht]
\includegraphics[width=30pc]{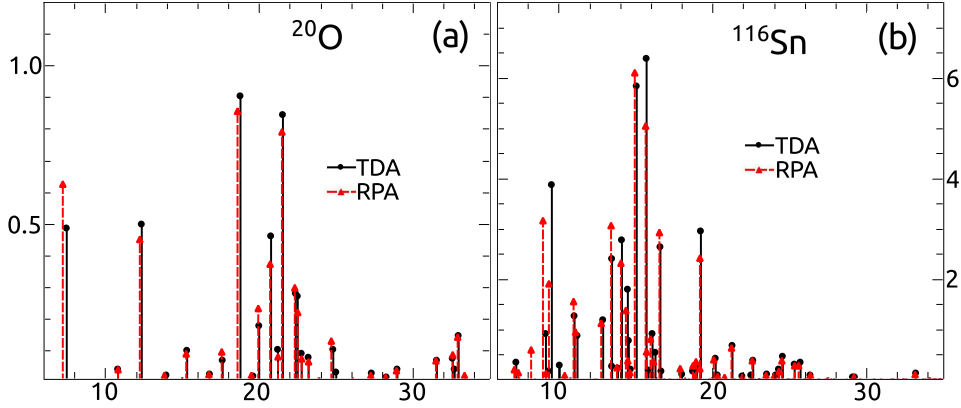}
\caption{  \label{fig6}     QTDA   versus  QRPA $E1$  spectra in $^{20}$O  (a) and  $^{116}$Sn (b).
} 
\end{figure}

Indeed, the  QTDA spectra, free of spurious admixtures,  are very similar to the corresponding standard QRPA spectra, apart from minor differences, especially  at low energy. This is illustrated in Figure \ref{fig6} for  $^{20}$O   and $^{116}$Sn.
  
\section{QTDA and QRPA E1 response} 
 We apply both QTDA and QRPA to the study of the $E1$ response. To this purpose we compute the cross section
\begin{eqnarray}
\sigma_{int}  =  \int_{E_0}^{E } \sigma (\omega) d\omega = \frac{16 \pi^3 e^2}{9\hbar c} \int_{E_0}^{E } \omega {\cal S}(E1, \omega) d\omega. 
 \label{sigma}
\end{eqnarray} 
where   ${\cal S}(E1, \omega)$ is the $E1$ strength function given by Eq. (\ref{Strength}). This reduced strength entering ${\cal S}(E1, \omega)$  
is evaluated by means of the modified $E1$ operator (\ref{E1effch}).
 
The cross section is simply proportional to  the classical energy weighted Thomas-Reiche-Kuhn (TRK) sum rule
\begin{eqnarray}
 S(TRK) = \sum_n \omega_n B_n (E1) = \frac{\hbar^2}{2 m } \frac{9}{4 \pi} \frac{N Z}{A} e^2. 
 \label{TRK}
\end{eqnarray}
We have in fact
\begin{eqnarray}
\sigma_{int}  =    \frac{16 \pi^3}{9\hbar c} S(TRK) = (2 \pi)^2 \frac{\hbar^2}{2m} \frac{e^2}{\hbar c}  \frac{N Z}{A} = 60 \frac{N Z}{A} (MeV mb). 
 \label{sigmaTRK}
\end{eqnarray}

For some specific nuclei we have also computed the isoscalar dipole transition strength using the operator 
\begin{eqnarray} 
{\cal M}_{IS} (\lambda =1, \mu) = 
\sum_{i=1}^A   \Bigl( r_i^{3} - \frac{5}{3} <r^2> \Bigr) Y_{1 \mu}(\hat{r_i}). \label{E1IS}
\end{eqnarray}
The second term in the bracket removes, partially, the CM contribution to the QRPA transitions strength. This term is unnecessary in the improved QTDA used here.

\begin{figure}[ht]
\includegraphics[width=30pc]{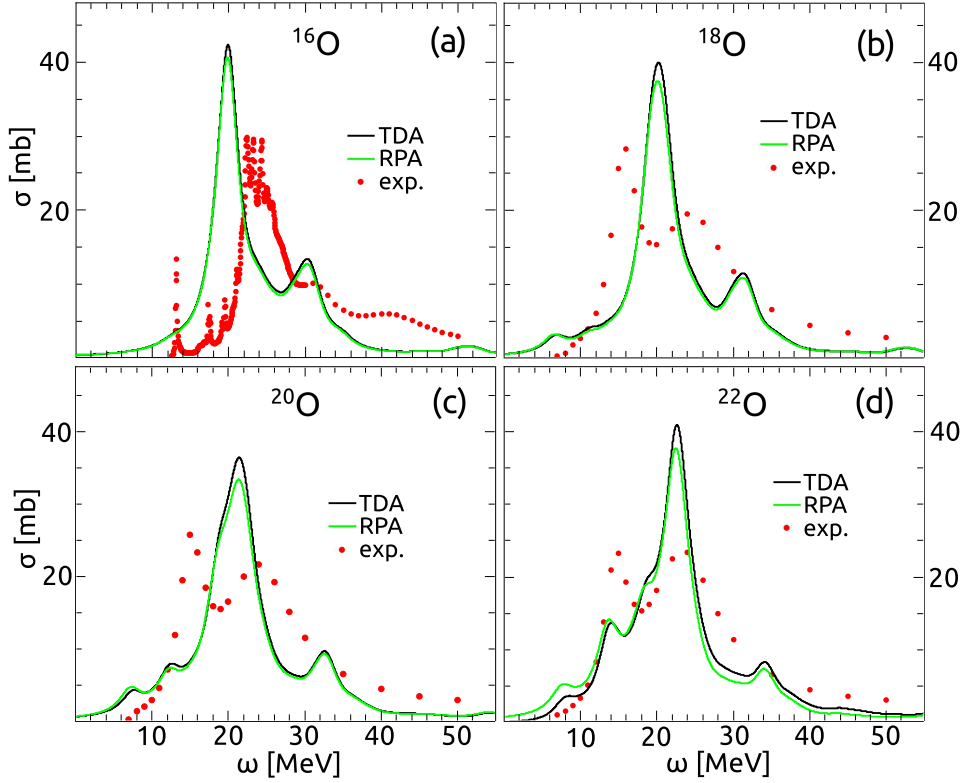}
\caption{\label{fig7}  Theoretical versus experimental $E1$ cross sections in $^{16-22}$O isotopes. The experimental data are taken from Ref. \cite{jendl} for $^{16}$O and from \cite{tendl,tendla} for the other isotopes. The theoretical cross sections were computed in  QTDA and QRPA.} 
\end{figure}

\begin{figure}[ht]
\includegraphics[width=30pc]{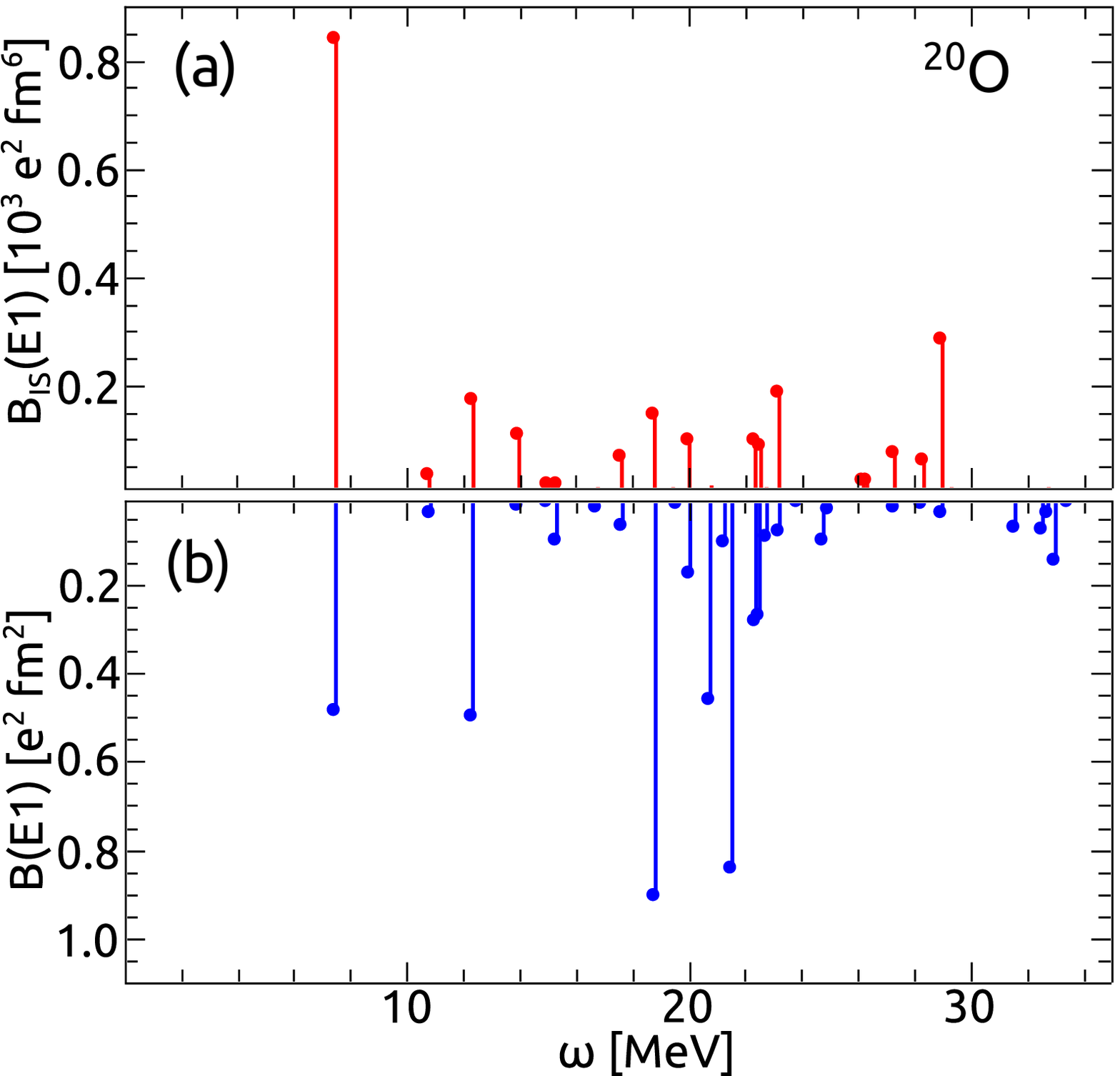}
\caption{\label{fig8}  Distribution of  isoscalar (a) and isovector (b) dipole reduced transition strengths  in $^{20}$O.
} 
\end{figure}

\subsection{Dipole response  in oxygen isotopes}

The cross section is computed by using  a Lorentzian of  width
 $\Delta= 3.5$ MeV for all oxygen isotopes.
 
Both QTDA and QRPA $E1$ cross sections are plotted in  Figures \ref{fig7}   for the doubly-closed shell $^{16}$O and the neutron-rich $^{18-22}$O isotopes.    The theoretical spectra are compared with the available data \cite{jendl,tendl,tendla}. 

One can hardly notice any difference between    the two cross sections.  Even their integrated values are close.
In fact, the  QTDA integrated cross section  overestimates the TRK sum rule by a factor $\sim 1.54$ in $^{16}$O and $\sim 1.48$ in $^{20}$O close to the corresponding QRPA factors  $\sim 1.43$ and $\sim 1.42$, respectively. The violation of the TRK sum rule in QRPA,  consistent with the results obtained in Ref. \cite{HergPap11}, is to be ascribed to momentum dependent components of the two-body potential.

The plots show important discrepancies between theory and experiments. In fact, the computed cross sections    compare only qualitatively   with the experimental data.  In the doubly magic $^{16}$O, the theoretical peak lies  $\sim 4$ MeV below the experimental one. In the open shell isotopes, the calculation yields one prominent peak just above 20 MeV and a smaller one $\sim$10 MeV above. Also the experimental cross sections exhibit  two peaks,     one at $\sim 17-18$ Mev and the other at $\sim 24-26$ MeV. The two peaks, however,    are of comparable height.

Consistently with experiments, the theoretical cross section below $\sim 15$ MeV is negligible in the $N=Z$  $^{16}$O, exhausting only  $1.5\%$ of the  classical TRK sum rule.  It   exhausts, instead,  a considerable fraction of the classical sum rule in the $N > Z$   isotopes. This fraction is $\sim 10 \%$ in $^{18}$O, $\sim 15 \%$ in $^{20}$O, $\sim 13 \%$ in $^{22}$O. These values are only slightly larger than the quantities deduced from the data \cite{Leiste}. These low-lying excitations were tentatively associated to the Pygmy resonance. Their collective character, however, has been questioned \cite{Paar07}. 

Just to gain  a closer insight into the structure and nature of the dipole spectra we plot in Figure \ref{fig8} the isovector and isoscalar dipole strength distributions  of $^{20}$O. We show only the  QTDA spectra since the ones obtained in QRPA are very similar. As pointed out already, the   QTDA states are completely free of spurious admixtures. Thus, the CM corrective terms   in the isoscalar (Eq. (\ref{E1IS})) and isovector (Eq. (\ref{E1effch})) operators  are irrelevant.

The low-lying isovector $E1$ spectrum is composed mainly of  two strong  peaks, one  at $\omega_{1^-} = 7.46$ MeV, almost at the neutron decay threshold ($7.608$ MeV), and another well above the neutron decay threshold at $\omega_{1^-} = 12.1$ MeV.  The strengths collected by these two $1^-$ states are $B(E1)= 0.49$ $ e^2 fm^2$ and 
$B(E1)= 0.50$ $ e^2 fm^2$  respectively.  

Experimentally,  two   $E1$ excitations    below the neutron threshold, in the $5-7$ MeV energy interval, were  observed in a virtual-photon scattering experiment \cite{tryg02,tryg03,Aumann08}. These two states, however, collect an almost negligible $E1$ integrated strength, $\sum B(E1) \sim 0.1$ $ e^2 fm^2$. 

Thus,  almost the total low-energy $E1$ strength lies above the neutron decay threshold \cite{Leiste,Aumann08}. 
We may therefore conclude, that the  theoretical low-lying $E1$ strength distribution is consistent with the experimental findings.
   
Figure \ref{fig8} shows that the lowest $1^-_1$ is strongly excited also by   the isoscalar operator (\ref{E1IS}). This is, actually, the strongest isoscalar dipole peak. The rest of the strength is distributed almost uniformly in the whole energy interval.

It may be of interest to have a quick look at the wave functions. The lowest $1^-$ state of energy $\omega_{1^-} = 7.46$ MeV is built almost entirely of neutron excitations. In fact the neutron component represents the $94\%$ of the state. The dominant    neutron quasi-particle components are  $\mid (1p_{3/2})_n (0d_{5/2})_n \rangle$ and  $\mid (1p_{3/2})_n (1 s_{1/2})_n \rangle$ which account for $43 \%$ and $34\%$, respectively, of the total wavefunction. 

These configurations describe excitations of valence neutrons   and, therefore,  emphasize the role of the neutron skin, suggesting strongly the pygmy nature of the low-lying transitions. The other strongly excited state at $\omega_{1^-} = 12.1$ MeV has a similar nature. It is in fact dominated by the neutron   configuration $\mid (0f_{7/2})_n (0 d_{5/2})_n \rangle$ with a weight of $66\%$. 

We have also evaluated the proton and neutron weights of the $1^-$ state of energy  $\omega_{1^-} = 21.54$ MeV, well in the region of the GDR. The neutron component of this state is $61\%$, still dominant but not overwhelming.  
Its  largest components are the neutron  configuration  $\mid (0d_{3/2})_n (0 p_{1/2})_n \rangle$  with a weight $53\%$ as well as the proton states $\mid (0d_{5/2})_p (0 p_{3/2})_p \rangle$ and $\mid (1s_{1/2})_p (0 p_{3/2})_p \rangle$ with weights $19\%$ and $13\%$, respectively. These configurations describe excitations of protons and neutrons from the core and, therefore,   qualify the peak as a member of the GDR.

\begin{figure}[ht]
\includegraphics[width=30pc]{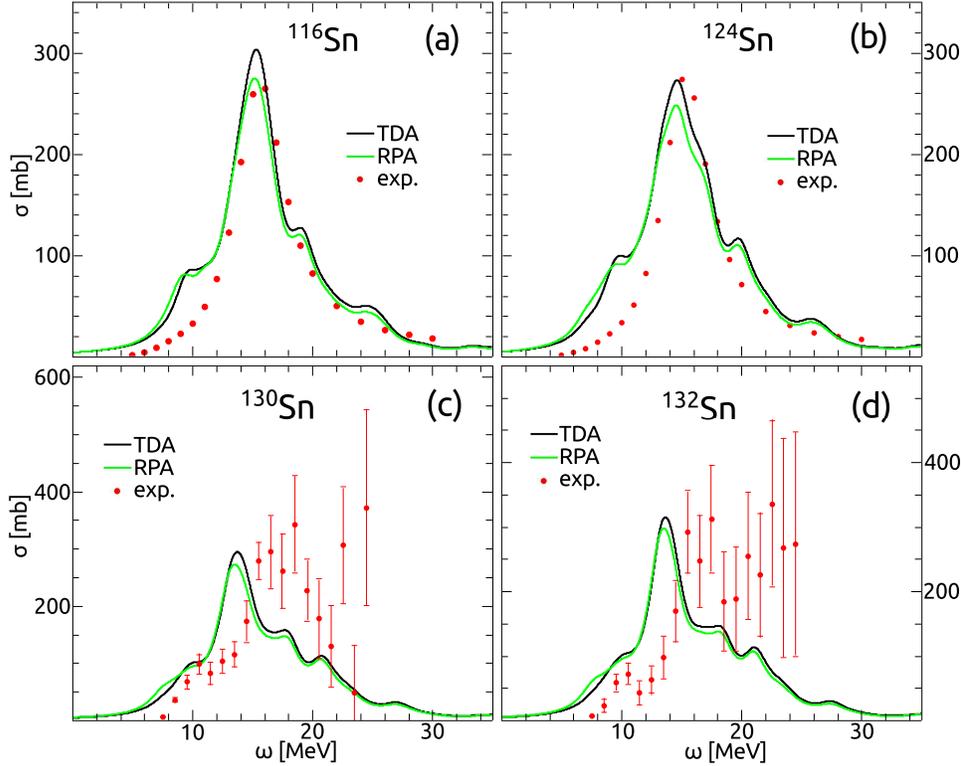}
\caption{ \label{fig9}  Theoretical versus experimental $E1$ cross sections in some tin isotopes. The theoretical cross sections were computed in  QTDA and QRPA.  The experimental data    are taken from \cite{Adrich05} for   $^{130,132}$Sn and  from \cite{tendl,tendla} for $^{116,124}$Sn} 
\end{figure}

\begin{figure}[ht]
\includegraphics[width=30pc]{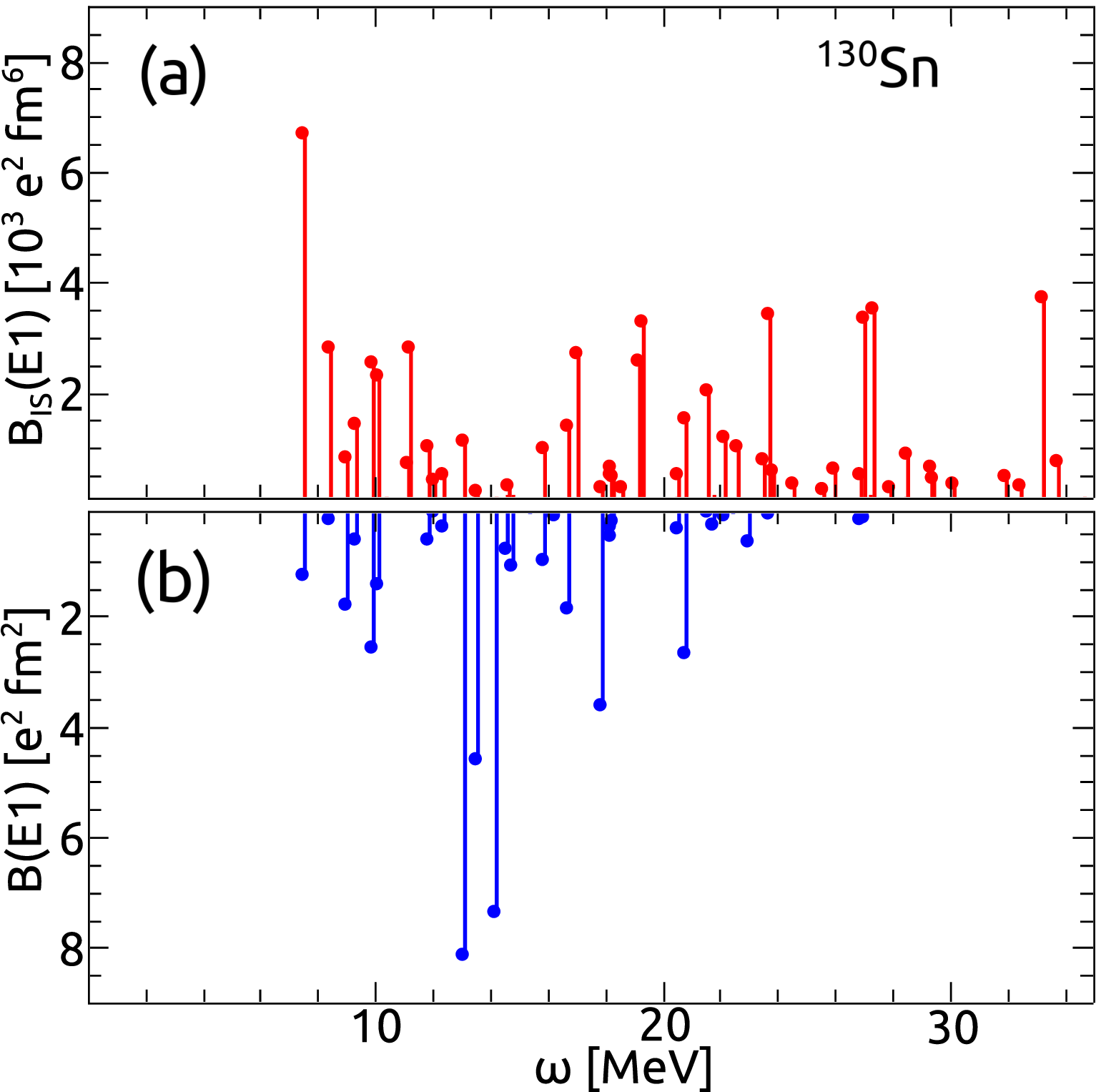}
\caption{\label{fig10}  Distribution of the   isoscalar (a) and isovector (b) dipole  reduced transition strengths  in $^{130}$Sn.
} 
\end{figure}

\subsection{Dipole response in Sn isotopes} 
QTDA and QRPA cross sections for some typical Sn isotopes are plotted in Figure \ref{fig9}. They were computed using a Lorentzian width $\Delta = 2.5$ MeV.

QTDA and QRPA came out to yield very similar strength distributions also in these isotopes.  In both approaches, the TRK sum rule is overestimated by a comparable amount. The QTDA enhancement factor is $\sim 1.54$ in both $^{130}$Sn and $^{132}$Sn isotopes, close to the QRPA corresponding values  $\sim 1.46$ and  $\sim 1.47$, respectively.
 
In $^{130,132}$Sn the computed cross section is   $\sim 2$ MeV below   the one measured through Coulomb excitation experiments \cite{Adrich05}. Apart from the shift, it evolves according to a law which is  fully  compatible with the data   in the low-energy sector and     with the values in the high energy sector characterized by large errors.

The agreement between theory and experiments is more  satisfactory in the   Sn isotopes far away from the neutron shell closure, where  the theoretical curves follow fairly closely the measured points and are peaked in the right position.  

The plots show clearly that a non negligible strength is concentrated at low-energy around the neutron decay threshold. The fraction of TRK sum rule exhausted by the low-lying states up to $\sim 8.5$ MeV is $\sim5\%$ in $^{116}$Sn, $\sim4\%$  in $^{124}$Sn, $\sim6\%$ in $^{130}$Sn, $\sim4\%$ in $^{132}$Sn. These fractions are consistent with the experimental ones, which are $7(3)\%$ and $4(3)\%$ in $^{130}$Sn and $^{132}$Sn, respectively \cite{Adrich05}.

In order to elucidate the nature of these low-lying excitations   we  plot for $^{130}$Sn  the strength distribution of   the QTDA  isovector and isoscalar dipole transitions. 

As shown in Figure \ref{fig10}, at low energy   we have  two $1^-$ states strongly excited by the isovector $E1$ operator, one at $\omega_{1^-} = 7.49$ MeV which carries a strength $B(E1) = 1.27$ $e^2 fm^2$ and the other at $\omega_{1^-} = 9.01$ MeV collecting a strength $B(E1) = 1.71$ $e^2 fm^2$. The summed strength up to $\omega \sim 9$ MeV is $\sum_n B_n(E1) = 3.35$ $e^2 fm^2$, close to the experimental value $\sum_n B^{(exp)}_n(E1) = 3.2$ $e^2 fm^2$ \cite{Adrich05}. Similarly, for $^{132}$Sn, the summed strength up to $\omega \sim 9$ MeV is $\sum_n B_n(E1) = 2.1$ $e^2 fm^2$ in agreement with the measured $\sum_n B^{(exp)}_n(E1) = 1.9$ $e^2 fm^2$. 

Let us now look at the wave functions in $^{130}$Sn. The $1^-$ excitation at $\omega_{1^-} = 7.49$ MeV has the properties of a pygmy mode. Indeed,  the neutron    component $\mid (2p_{3/2})_n (2 s_{1/2})_n \rangle$ accounts for $85\%$ of this state and the total neutron weight is $95\%$. 

The state at $\omega_{1^-} = 9.01$ MeV has a more complex and collective structure. Neutrons and protons contribute with a comparable weight, $49\%$ and $51\%$ respectively. The dominant proton components are  $\mid (1d_{5/2})_p (1p_{3/2})_p \rangle$ and $\mid (1d_{3/2})_p (1p_{1/2})_p \rangle$ with respective weights $22\%$ and $23\%$. The dominant neutron states  are $\mid (2p_{3/2})_n (1 d_{3/2})_n \rangle$ and $\mid (2p_{1/2})_n (1 d_{3/2})_n \rangle$ with weights $28\%$ and $14\%$, respectively.
 
The $1^-$ state of energy  $\omega_{1^-} = 13.13$ MeV belonging to the group of GDR peaks has a collective nature with a neutron dominance ($68\%$). Its largest proton component is    $\mid (0h_{11/2})_p (0 g_{9/2})_p \rangle$  with a weight $23\%$. The dominant neutron configurations are    $\mid (1f_{7/2})_n (0 g_{7/2})_n \rangle$ and $\mid (0i_{13/2})_n (0 h_{11/2})_n\rangle$ with weights $27\%$ and $15\%$, respectively.  

\begin{figure}[ht]
\includegraphics[width=30pc]{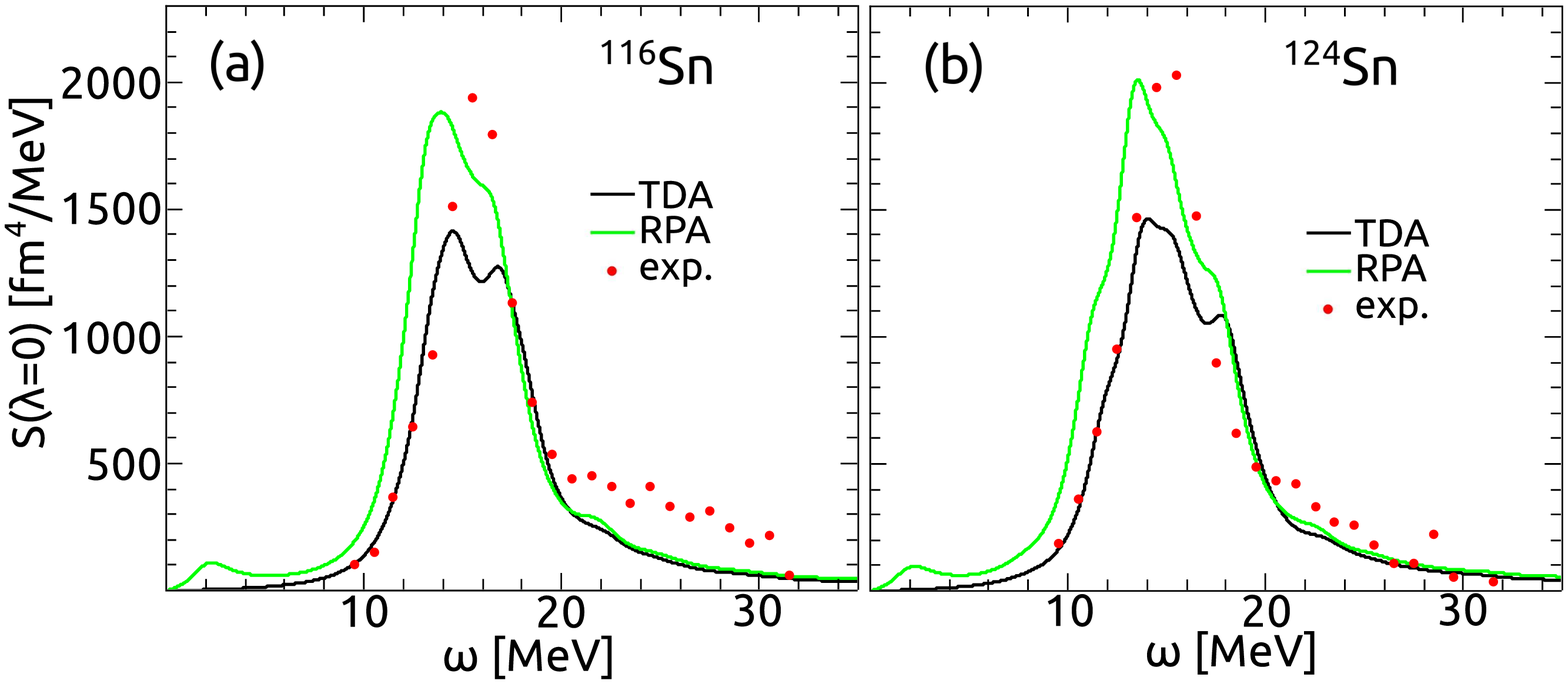}
 \caption{ \label{fig11}  Theoretical versus experimental monopole strength functions in $^{116}$Sn (a) and $^{124}$Sn (b).  The theoretical strength functions were computed in  QTDA and QRPA. The experimental data  are taken   from \cite{Li07,Li10}} 
\end{figure}
 
 \begin{figure}[ht]
\includegraphics[width=30pc]{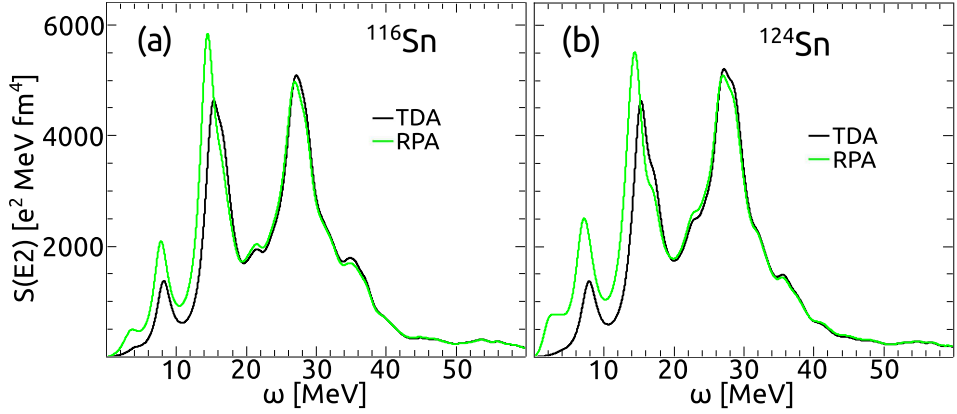} 
\caption{ \label{fig12}  QRPA versus QTDA $E2$ strength function distribution in $^{116}$Sn (a) and $^{124}$Sn (b) } 
\end{figure}

Figure \ref{fig10} shows that the isoscalar dipole spectrum not only overlaps with the isovector one but covers a larger region. Indeed, the isoscalar probe excites strongly the states in the $\sim 3 \hbar \omega$ region describing a compressional mode as well as the ones in the low-energy region associated to the pygmy resonance.  

The coexistence of an isovector and and isoscalar spectrum at low energy is consistent with a recent experiment \cite{Endres10}. This experiment has detected  a very dense spectrum in the $\sim 5.5- 8.5$ MeV interval, composed of two groups of levels. One group, observed in  $(\gamma, \gamma')$ reaction, is of isovector nature, the other, observed in $(\alpha, \alpha'\gamma)$, is isoscalar. All the isovector $E1$ transitions are very weak, of the order $B(E1) \sim 10^{-3} - 10^{-2}$ $e^2 fm^2$ and the integrated strength
is $\sum_n B_n(E1) \sim 0.5$ $e^2 fm^2$.  Our calculation  is unable to reproduce this highly  fragmented strength. In order to obtain such a rich spectrum it is necessary to go beyond QRPA by resorting for instance to QPM \cite{Tzon08} or the relativistic time blocking approximation  (RTBA) \cite{Litv0}.  

\begin{figure}[ht]
\includegraphics[width=30pc]{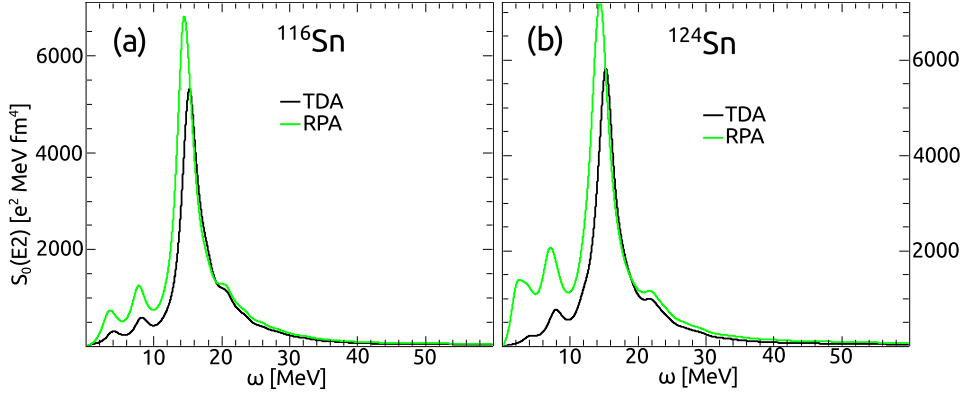} 
\caption{ \label{fig13}  QRPA versus QTDA isoscalar  strength function $S_0(E2)$  in $^{116}$Sn (a) and $^{124}$Sn (b) } 
\end{figure}

\begin{figure}[ht]
\includegraphics[width=30pc]{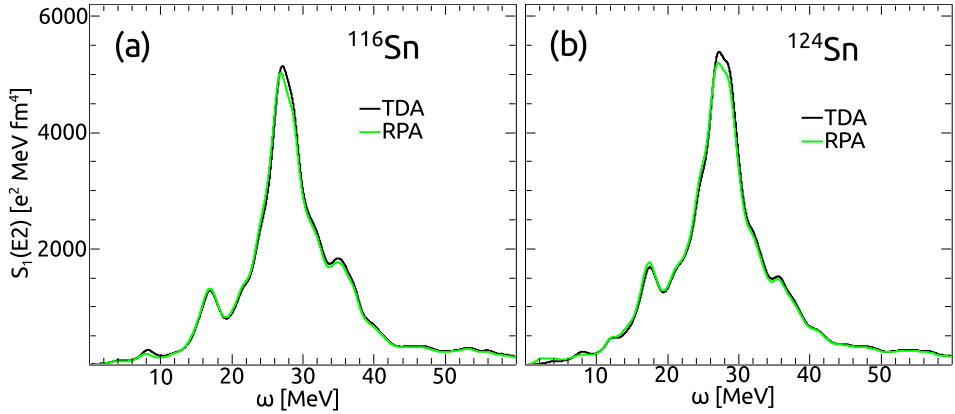} 
\caption{ \label{fig14}  QRPA versus QTDA isovector   strength function $S_1(E2)$  in $^{116}$Sn (a) and $^{124}$Sn (b) } 
\end{figure}
 
\section{Monopole and quadrupole responses}
As mentioned already, the quasi-particle energies within a major shell are too far from the empirical ones to allow a detailed study of the positive parity spectra. Thus,  we will give only a brief account of the global properties of the monopole and quadrupole responses. 

To this purpose we have computed the strength functions (\ref{Strength}) for  $^{116}$Sn and $^{124}$Sn  using a Lorentzian of width $\Delta =2.5$ MeV. The $E2$ transition strengths were computed by using the operator given by the standard formula (\ref{MElamb}) with the bare charges $e_p=1$ and $e_n=0$. For the monopole transitions we used the operator
\begin{eqnarray}  
{\cal M} (\lambda =0) =  
\sum_{i=1}^A    r_i^{2} Y_{00}(\hat{r_i}). \label{E0}
\end{eqnarray}
As shown  in Figures \ref{fig11}, the QRPA monopole strength function describes fairly well the experimental trend \cite{Li07,Li10} and is in fair agreement with the response evaluated within a density functional approach \cite{Ligan12}. The  QTDA strength, instead, follows the evolution of the data but has a lower peak.

Unlike the dipole case, appreciable differences between QTDA and  QRPA responses appear in the low energy sector. These difference are also clearly visible in   the $E2$ strength distributions shown in Figure  \ref{fig12}. Apparently, the QRPA ground state correlations are more effective in the low-energy sector composed predominantly of    isoscalar transitions. No differences are noticed in the high energy part of the spectrum characterized mainly by  isovector transitions. The above statements are explicitly confirmed by the plots of isoscalar and isovector $E2$ strength distributions shown in Figures \ref{fig13} and \ref{fig14}, respectively.

\section{Concluding remarks}
The present calculation has  confirmed that the HFB equations, using only a realistic potential $V_{lowk}$ deduced from the bare NN interaction, generate completely unrealistic quasi-particle spectra.  These become compatible with experiments only  if one adds to $V_{lowk}$ a density dependent two-body potential $V_\rho$ simulating a three-body contact force, first adopted in   Ref. \cite{HergPap11}. 

Although the improvement promoted by $V_\rho$ is impressive, appreciable discrepancies between theory and experiments remain.  The spin-orbit term added here improves the spectra by enhancing the spin-orbit splitting but is not able to reduce sufficiently the distance between the quasi-particle configurations within  major shells as it would be necessary in order to fill the gap with the empirical data. This point deserves further investigation.

The mentioned limitation, however, is expected to  affect  the low-energy spectra and to have a marginal impact of the multipole responses. We adopted  both  QTDA and QRPA to study mainly the dipole excitations. We discussed briefly also the monopole and quadrupole transitions.
  
Appreciable differences between the QTDA and QRPA are noticeable only in the low-energy sectors of the monopole and quadrupole spectra, while the  dipole responses are practically identical in the two approaches. 

These results indicate that  the QRPA ground state correlations affect only the isoscalar modes. In particular,    
they seem to improve the   description of the monopole  strength distribution as our comparative analysis has suggested.

The dipole cross sections, whether computed in  QTDA or QRPA, are in fair agreement with experiments  in tin isotopes apart from a $\sim 2$ MeV displacement between   theoretical and experimental peaks  observed in few isotopes.   The agreement with experiments is only qualitative in oxygen isotopes.  

The calculations yield in nuclei with neutron excess an appreciable low-energy dipole strength comparable to the one measured in recent experiments \cite{Adrich05}. The low-lying states are shown to be excited not only by isovector but also by isoscalar probes, consistently with recent experimental findings \cite{Endres10}. On the other hand, the high fragmentation observed experimentally is far from being reproduced within QTDA or QRPA. 

In order to be able to describe such a rich spectrum of weakly excited levels it is necessary to go beyond the harmonic approximation.   One approach of this nature is the  EMPM which proved to be able to  induce a very pronounced fragmentation at low energy in the neutron rich  $^{208}$Pb \cite{Biancoa}. The method has been already formulated in the language of quasi-particles \cite{LoBia} and is therefore suited to the study of complex spectra in open shell nuclei. In fact, a HFB self-consistent EMPM calculation of the dipole response for the nuclei investigated here    is under way.

In summary,  it is possible to improve drastically the QTDA or QRPA description of the multipole nuclear strength distributions starting from a  $V_{lowk}$ potential deduced from a realistic $NN$ interaction, if one adds to $V_{lowk}$   a density dependent plus a spin-orbit corrective terms. This is just an ad hoc prescription and, therefore, not satisfactory on theoretical ground. 

The final goal is to explore if comparable descriptions are obtained by the  exclusive use of realistic potentials, like the ones deduced from chiral effective field theory \cite{EntMac03},  which include $NN$ interactions with accompanying three-body terms. The latter terms have been shown to play an important role  in nuclear structure \cite{Ham13}.  
The phenomenological density dependent term, so crucial for reducing the gap between theoretical and experimental spectra, may be one of the elements to be taken into account in the fine tuning parametrization of the chiral three-body force. 
The spin-orbit corrective term might be one of the inspiring elements for a revision of the parameters of the $NN$ tensor forces which  is known to affect strongly      the spin-orbit splitting.  This aspect was emphasized recently in a  mean field  model  using a meson exchange tensor potential \cite{Ots06} and  a HF+BCS approach using a Skyrme force \cite{Colo07}.

\ack
One of the authors (F. Knapp) thanks for partial financial support the Istituto Nazionale di Fisica Nucleare (INFN) as well as   the Czech Science Foundation (Grant No. 13-07117S) and Charles University in Prague (project UNCE 204020/2012)


\section*{References}
\begin{thebibliography}{99}

  
\bibitem{Doba84}  Dobaczewski J,  Flocard H and Treiner J 1984  {\it Nucl. Phys. A}  {\bf  422} 103 

\bibitem{Doba96}  Dobaczewski J,   Nazarewicz W,   Werner T R,  Berger J F,  Chinn C R  and  Decharge J 1996
{\it Phys. Rev. C} {\bf 53} 2809 

\bibitem{engel99}  Engel J,  Bender M,  Dobaczewski J,  Nazarewicz W and  Surman R 1999 {\it Phys. Rev. C} {\bf 60} 014302

\bibitem{Khan02}  Khan E,  Sandulescu N,  Grasso M and  Giai Nguyen Van 2002
{\it Phys. Rev. C} {\bf 66} 024309 

\bibitem{Tera05}  Terasaki J,  Engel J,  Bender M,  Dobaczewski J,  Nazarewicz W and  Stoitsov M 2005 {\it Phys. Rev. C} {\bf 71} 034310 

 \bibitem{Yosh08} Yoshida K and Van Giai N 2008 {\it Phys. Rev.} {\bf 78} 064316 
  
  \bibitem{Miz09} Mizuyama K,  Matsuo M and  Serizawa Y 2009 {\it Phys. Rev. C} {\bf 79} 
   024313 
 
\bibitem{Losa}  Losa C,  Pastore A,  Døssing T,  Vigezzi E and  Broglia R A 2010 {\it Phys. Rev. C}  {\bf 81} 064307 

\bibitem{Giamb03} Giambrone G,  Scheit S,  Barranco F, Bortignon P F,  Colo G,  Sarchi D,  Vigezzi E 2003 {\it Nucl. Phys. A}  {\bf  726} 3 

\bibitem{Peru08}  Peru S and  Goutte H 2008 {\it Phys. Rev. C}  {\bf 77} 044313   

\bibitem{Paar03} Paar N,  Ring P,  Niksic  T and  Vretenar D 2003 {\it Phys. Rev. C} {\bf 67} 034312

\bibitem{Bend03}  Bender M,  Heenen P-H and  Reinhard P-G 2003 {\it Rev. Mod. Phys.} {\bf 75} 121 

\bibitem{Vret05}   Vretenar D,  Afanasjev A V,  Lalazissis G A and  Ring P 2005 {\it Phys. Rep.} {\bf 409} 101

\bibitem{Herg09}  Hergert H and  Roth R 2009 {\it Phys. Rev. C} {\bf 80} 024312 

\bibitem{HergPap11} Hergert H,  Papakonstantinou P and Roth R 2011 {\it Phys. Rev. C} {\bf 83} 064317 

\bibitem{RiSchu} Ring P and  Schuck P 1980 {\it The Nuclear Many-Body Problem}  (Berlin: Springer)

\bibitem{UCOM} Feldmeier H,  Neff T,  Roth R and  Schnack J 1998 {\it Nucl. Phys. A} {\bf 632} 61 

\bibitem{argonne} Wiringa R B,  Stoks V G J and  Schiavilla R 1995 {\it Phys. Rev. C} {\bf 51} 38 

\bibitem{SRG} Bogner S K,  Furnstahl R J and  Schwenk A 2010 {\it Prog. Part. Nucl. Phys.} {\bf 65} 94 

\bibitem{Mcl}  Machleidt R 2001  {\it Phys. Rev. C} {\bf 63}  024001 

\bibitem{Bogner01}  Bogner S K,  Kuo T T S and   Coraggio L 2001 {\it Nucl. Phys. A} {\bf  684} 432c

\bibitem{Bogner03} S. K. Bogner S K,  Kuo T T S and  Schwenk A 2003 {\it Phys. Rep.} {\bf 386} 1 

\bibitem{covrev} For references Coraggio L, Covello A, Gargano A, Itaco N and  Kuo T T S 2009 {\it Prog. Part. Nucl. Phys.} {\bf 62} 135 


\bibitem{Bianco}  Bianco D,   Knapp F,  Lo Iudice N, Andreozzi F, and  Porrino A 2012
{\it Phys. Rev. C}  {\bf 85} 014313 

\bibitem{Paar07}  Paar N,  Vretenar D,  Khan E and  Col$\grave{o}$ G 2007 {\it Rep. Prog. Phys.} {\bf 70} 691  for review and early references

\bibitem{LoI12}  Lo Iudice N,  Ponomarev V Yu ,  Stoyanov Ch,  Sushkov A V and  Voronov V V 2012
{\it J. Phys. G: Nucl. Part. Phys. }{\bf 39}  043101

\bibitem{Biancoa}   Bianco D,   Knapp F,  Lo Iudice N,  Andreozzi F,  Porrino A and  Vesely P 2012
{\it Phys. Rev. C}  {\bf 86} 044327 
%\bibitem{Jens} Hjorth-Jensen M,  Kuo T T S and Osnes E 1995 {\it Physics Reports}  {\bf 261} 125 

\bibitem{Leiste}  Leistenschneider A {\it et al.} 2001 {\it Phys. Rev. Lett.} {\bf 86} 5442 

\bibitem{tryg02}  Tryggestad  E {\it et al.} 2002 {\it Phys. Lett. B}  {\bf 541} 52 

\bibitem{tryg03}  Tryggestad E {\it et al.} 2003 {\it Phys. Rev. C} {\bf 67} 064309 

\bibitem{Saga}  Sagawa H and  Suzuki T 1999 {\it Phys. Rev. C} {\bf 59} 3116 

\bibitem{CoBo01} Col$\grave{o}$ G and Bortignon P F 2001 {\it Nucl. Phys. A} {\bf 696} 427 

\bibitem{Adrich05} Adrich P {\it et al.} 2005 {\it Phys. Rev. Lett.} {\bf 95} 132501 

\bibitem{Oze07}  Ozela B,  Enders J, von Neumann-Cosel P,  Poltoratska I, Richter A,  Savran D,
 Volz S,  Zilges A 2007 {\it Nucl. Phys. A} {\bf 788} 385c
 
\bibitem{Endres10}  Endres J {\it et al.} 2010 {\it Phys. Rev. Lett.} {\bf 105} 212503 

\bibitem{Sar04}  Sarchi D,  Bortignon P F, Col$\grave{o}$ G 2004 {\it Phys. Lett. B} {\bf 601} 27

\bibitem{Tzon08}  Tsoneva N and  Lenske H 2008 {\it Phys. Rev. C} {\bf 77}  024321

\bibitem{Avdee11} Avdeenkov A,  Goriely S, Kamerdzhiev S,  Krewald S 2011 {\it Phys. Rev. C} {\bf 83}  064316

\bibitem{Klim07} Klimkiewicz A  {\it et al.} 2007   {\it Phys. Rev. C} {\bf 76} 051603(R) 

\bibitem{Daut11} Daoutidis I and  Ring P 2011 {\it Phys. Rev. C} {\bf 83}    044303

\bibitem{Aumann08} Aumann T 2008 {\it Nucl. Phys. A} {\bf 805 } 198c for review and references

\bibitem{Enge}  Engeland T,  Hjorth-Jensen M and Jansen G R  {\it  Computational environment for nuclear structure} (CENS),  
http://folk.uio.no/mhjensen/cp/software.html.

\bibitem{Waroq}  Waroquier M,  Heyde K and  Vincx H 1976 {\it Phys. Rev. C} {\bf 13} 1664 

\bibitem{Gunter} Gunther A,  Roth R, Hergert H and  Reinhardt S 2010 {\it Phys. Rev. C} {\bf 82} 024319

\bibitem{Isak02}  Isakov V I,  Erokhina K I,  Mach H, Sanchez-Vega M, and Fogelberg B 2002 {\it Eur. Phys. J. A} {\bf 14} 29 

\bibitem{jendl}JENDL Photonuclear Data File 2004 (JENDL/PD-2004)

\bibitem{tendl} "TENDL-2012: TALYS-based evaluated nuclear data library", A.J.
Koning, D. Rochman, S. van der Marck, J. Kopecky, J. Ch. Sublet, S.
Pomp, H. Sjostrand, R. Forrest, E. Bauge and H. Henriksson,
www.talys.eu/tendl-2012.html

\bibitem{tendla}  Koning A J and  Rochman D 2012 {\it Nuclear Data Sheets} {\bf 113}  2841

\bibitem{Litv0} Litvinova E,  Ring P and  Tselyaev V 2007 {\it Phys. Rev. C} {\bf 75} 064308 

 
\bibitem{Li07} Li T {\it et al.} 2007 {\it Phys. Rev. Lett.} {\bf 99} 162503 

\bibitem{Li10} Li T {\it et al.} 2010 {\it Phys. Rev. C} {\bf 81 } 034309 

\bibitem{Ligan12} Li-Gang Cao   Sagawa H and  Col$\grave{o}$ G 2012 {\it Phys. Rev. C} {\bf 86 } 054313

\bibitem{LoBia} Lo Iudice N, Bianco D, Knapp F Andreozzi F, Porrino A,  Vesely P 2012 {\it J. Phys.: Conf. Ser.} {\bf 366} 012031

\bibitem{EntMac03} Entem D R and Machleidt R 2003 {\it Phys. Rev. C} {\bf 68} 041001(R) 

\bibitem{Ham13} Hammer H W, Nogga A and  Schwenk A 2013 {\it Rev. Mod. Phys.} {\bf 85} 197

\bibitem{Ots06} Otsuka T Matsuo T and  Abe1 D 2006 {\it Phys. Rev. Lett.} {\bf 97} 162501

\bibitem{Colo07} Col$\grave{o}$ G, Sagawa H, Fracasso S, Bortignon P F  2007 {\it Phys. Lett. B} {\bf 646} 227

%\bibitem{Paar03}  Paar N,  Ring P,  Niksic T, Vretenar D 2003 {\it Phys. Rev. C} {\bf 67} 034312

\end{thebibliography}
\end{document}